# Fuzzy Logic Approach for Threat Prioritization in Agile Security Framework using DREAD Model

Sonia[1], Archana Singhal[2] and Hema Banati[3]

[1]Department of Computer Science, University of Delhi, Research Scholar
Delhi, India

[2] Department of Computer Science, University of Delhi, IP College for Women
Delhi, India

[3] Department of Computer Science, University of Delhi, Dyal Singh College
Delhi, India

## Abstract

For a qualitative system sound security practices must be a crucial part throughout the entire software lifecycle. Furthermore, agile software development has paved the way for overcoming the problems faced by developers during traditional development process. In the given paper we are using an Agile Security Framework that is compatible with practices of agile processes and inherit in it the benefits of security engineering activities in the form of risk assessment and threat prioritization. One of the most popular techniques to deal with ever growing risks associated with security threats is DREAD model. It is used for rating risk of threats identified in the abuser stories. In this model threats needs to be defined by sharp cutoffs. However, such precise distribution is not suitable for risk categorization as risks are vague in nature and deals with high level of uncertainty. In view of these risk factors, our paper proposes a novel fuzzy approach using DREAD model for computing risk level that ensures better evaluation of imprecise concepts. Thus it provides the capacity to include subjectivity and uncertainty during risk ranking. A case study has been presented to illustrate and compare the proposed approach with the existing one using Matlab.

***Keywords:*** *Fuzzy logic, Agile Security Framework, Risk ranking, DREAD Model, Fuzzy Inference System.*

## 1. Introduction

System security is one of the most significant issues in today's software society. Several security threats arise during software development process. Software failures, due to various vulnerabilities present in software, suggest essential presence of security in every phase of software development method. It is observed that developers prefer to adopt agile methods for software development as compared to traditional development processes as agile methods provides software development at fast pace and ever-changing. Agile processes are gaining popularity mainly during development of web applications as prevailing conditions recognize that here changes are inevitable and security risks are more prominent. Agile methodology a lightweight, iterative approach understands the need of current time, encouraging changes in requirements at any stage in software development lifecycle. However, integration of security measures with agile processes imposes several constraints making it imperative that these integration problems are analyzed carefully.

Sonia et al. [1] proposed an Agile Security Framework (ASF), an iterative framework presenting step by step guidance for applying security techniques wherein agility is maintained by providing flexibility in implementing changes at any stage. In the given framework, a hybrid technique has been suggested that combines abuser stories with attack trees to map security requirements. Moreover, a security framework that helps in categorizing security requirements has also been provided in ASF for different iterations. This framework continues to evolve and to be applied in better way integrating security activities wherever possible to get a complete secure system. There, in phase 2 of threat modeling and designing we focus just on one aspect of it that is, risk assessment and prioritization. This step deals with assigning risk rating. It gives high risk value, if threat poses significant risk and need to be addressed immediately.

Microsoft's DREAD model is a popular approach for computing risk level of threats but it allows only crisp values [7]. Virtually every risk element can be characterized using two metrics, "Low, Medium, and High," or through "Ordinal Ranking." [2]. Therefore most appropriate approach for defining risk level is using fuzzy logic. In this truth or validity of any statement becomes its degree of belongingness or membership. This degree corresponds to a v alue to which an object is similar or compatible with the concept represented by fuzzy set. Truthfulness of a s tatement can be of various degrees which ranges from completely true, to partially true and then to completely false [2]. Moreover, fuzzy logic has linguistic values taken as words which can represent natural language for human reasoning during fuzzy rules construction. Thus, our approach based on fuzzy logic, can easily deals with ambiguities and uncertainties posed by imprecise concepts. After threats are prioritized based on risk rating, appropriate action is taken to manage that particular risk. Ways listed under risk management includes risk acceptance, risk transference, risk removal and the last one is to mitigate risk by countermeasures [5]. Various steps for computing risk rating with overall process for threat prioritization using fuzzy approach is illustrated in Fig. 1.

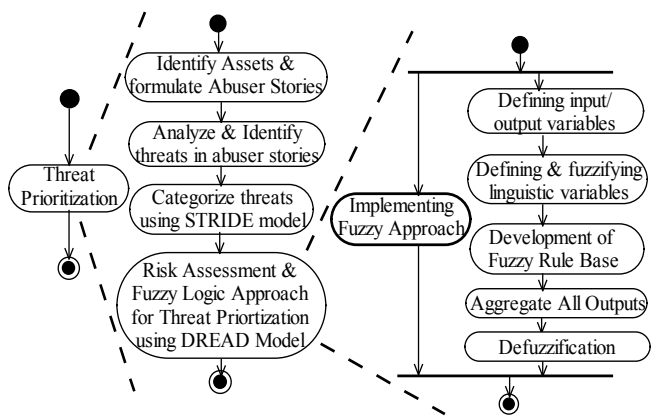

Fig. 1 Overall Threat Prioritization Process

The remainder of this paper is organized as follows. Section 2 presents the related work in the concerned field. Section 3 provides a brief overview required to understand our approach. Section 4 describes our proposed fuzzy approach. Application of our approach with a b rief case study represented in Section 5, and Section 6 presents conclusion and future work.

## 2. Overview of Existing Literature

Various publications presents security guidelines for software development and among them several proposals had delivered a quality work for implementing security in agile software development. Mikko Siponen has explained that how security techniques can be added seamlessly to agile software development using an agile method, named Feature Driven Development [3]. Gustav Bostrom et al. had proposed a method of extending extreme programming to support security requirements engineering [4]. In [5],

Suvda Myagmar has described threat modeling as a basis of security requirements. K. Ram Mohan Rao has done improvement in security by giving a web application risk assessment technique called threat risk modeling [6]. Microsoft used a DREAD model for assessing each threats relative risk [7]. Supreeth Venkataraman has prioritized threats using k/m algebra [8]. Several researchers have put forward an idea of implementing fuzzy logic for threat modeling. Fuzzy set theory was established by Lofti Zadeh in 1965 and he later provides some advance techniques in fuzzy approach for dealing with complex systems [9, 10]. Klir and Kosko further contributed by giving uncertainty and fuzzy rule base concept. They concluded that that probabilities for mutually exclusive events cannot add up to more than 1, but their fuzzy values can be like that [11, 12]. A.S. Sodiya, et al. have presented a f uzzy logic technique used to identify potential threats to computer based systems [13]. Other researchers have also focused on applying risk analysis in various applications using fuzzy logic [14]. But till now there is no study which provides fuzzy approach for threat risk ranking using DREAD model. Keeping these points in mind, we are proposing novel approach which will provide risk rating using fuzzy inference system.

## 3. Background

### 3.1 Agile Security Framework

To understand the amendments we have done in ASF it is necessary to first get the basic knowledge about the given framework [1]. As indicated by its name that, ASF is an iterative framework addressing security at various stages of its security development lifecycle to get a complete secure system. Here agility is maintained by providing flexibility in implementing changes at any stage. ASD is adaptive in nature, supporting continuous changes and produces working software at the end of each iteration. This framework is developed after keeping in mind all this aspects. As dealing with lightweight approach, even threat modeling process in this framework has time constraints, therefore it includes only current stories. It minimizes the development time and provides a w ay of keeping agile processes and security practices over a s ingle platform. This framework also suggests some key points which assist in providing iterative development to it. One of the key points describes that during first iteration, a security framework for categorizing abuser stories (which identify threats) is to be developed, which helps developer in understanding which story is to be implemented in which iteration. Another point discusses that initial iteration starts with release planning defining scope of whole project involving security requirements of complete system. ASF comprises of various phases from which phase 1 a nd 2 briefly describe some steps of Security analysis and Threat Modeling which will serve as a baseline for our approach. All 5 phases of ASF are explained in brief below

**Phase 1 - Security requirements analysis & planning:**

The main purpose of this phase is to identify critical assets and then formulate abuser stories describing undesired behavior of the system. Various steps included in this phase are as follows

*A. Critical asset identification & formulation of abuser stories*: Assets are resources that must be protected by system from an unauthorized user or an attacker. Initially developer identifies critical assets and important security features required by customers for developing abuser stories. After that developer will address some abuser stories and security user stories with constraints. These stories will illustrate how existing protective measures could be neglected or where a lack of such protection exists.

*B. Analyzing and Identifying threats:* Based on determined assets, security user stories and abuser stories, developers then determines potential attacks or threats by which adversary might try to affect an asset. After enumeration and identification, these threats can be categorized into six categories based on their effect using Microsoft's STRIDE model [7].

**Phase 2 - Threat Modeling and Designing:**

Given phase describes Threat Modeling and Designing, which act as foundation for specification of security requirements. Security design process includes threat analysis, various techniques to manage and mitigate risks and finally translate security requirements into reality. Now we will describe main stages comes under threat modeling and designing

*A. Risk Assessment and Threat Prioritization*: Each threat also has risk associated with it. For a threat identified in abuser stories, risk assessment is performed to map each threat either into mitigation mechanism or an assumption that it is not worth worrying about [4]. To formalize this process a list of priority with overall rating of threats is generated. We can compute the risk to an asset in conventional manner using a formula [15].

$$Risk = Impact * Probability$$

It uses probability based quantitative method. It has a drawback that estimating impact and probability is frequently difficult and team members usually will not agree on rating system giving equal distribution of the assets. To resolve this, Microsoft uses a DREAD model to compute a risk value for prioritization of threats. This model calculates security risks as an average of numeric values, assigned to each of the following categories.

- **D**amage potential: Rates the extent of the damage if attack succeeded.
- **R**eproducibility: Ranks how often an effort to reproduce an attack works.
- **E**xploitability: Estimate the value for the effort needed to exploit the threat.
- **A**ffected users: Estimate fraction of installations affected if an exploit widely available.
- **D**iscoverability: Measures that how easy is it to discover the vulnerability by an attacker.

Although this method is extremely popular till present day but it a llows exactly defined distributions or crisp boundaries. In this method, it is not possible to determine degree of belongingness for a particular threat risk. Therefore most appropriate approach for defining risk level is to express it in terms of linguistic variables and membership function used in fuzzy logic. Here membership grades are often represented by real numbers varying in closed interval between 0 and 1. Thus, in the given paper we are trying to improve the method suggested in ASF by giving a fuzzy approach for threat prioritization. Experimental Results of our extension suggest that it helps developers in dealing with imprecise data, ambiguities and uncertainties more efficiently as compared to existing one.

*B. Requirements Elicitation using a Hybrid Technique*

After threat prioritization, a Hybrid Technique for requirements elicitation is explained in ASF. The purpose of this technique is to map the threats identified for mitigation into security requirements using agile methodology. It achieves this by combining the strengths of abuser stories and attack trees. This phase ends with designing security requirements.

**Phase 3- Secure code Implementation:** Here they have used Test Driven Development where series of unit test are included for security stories of current sprint implementation.

Table 1: Linguistic Variables and their ranges

| S. No. | Linguistic Variables | Linguistic Value & Range | | Linguistic Value & Range | | Linguistic Value & Range | | Linguistic Value & Range | | Linguistic Value & Range | |
|---|---|---|---|---|---|---|---|---|---|---|---|
| 1 | Damage Potential (DP) | Negligible | | Slight | | Moderate | | Almost | | Catastrophic | |
| | | 0-2 | | 1-4 | | 3-6 | | 5-8 | | 7-10 | |
| 2 | Reproducibility (R) | Probably | | Likelihood | | Satisfiable | | Critical | | Vital | |
| | | 0-2.5 | | 1.5-4 | | 3.5-6 | | 5.5-8 | | 7.5-10 | |
| 3 | Exploitability (E) | Least | | Slight | | Moderate | | Almost | | Extreme | |
| | | 0-3 | | 2-5 | | 4-7 | | 6-9 | | 8-10 | |
| 4 | Affected users (AU) | Noticeable | | Satisfactory | | Average | | Disturbing | | Unbearable | |
| | | 0-2 | | 1-4 | | 3-6 | | 5-8 | | 7-10 | |
| 5 | Discoverability (D) | Least | | Slight | | Moderate | | Almost | | Extreme | |
| | | 0-2 | | 1.5-5 | | 3.5-7 | | 5.5-9 | | 7.5-10 | |
| 6 | Fuzzy Risk Level (Output Variable) | Very Low | Low | Somewhat Low (S_WLow) | Medium | Somewhat High (S_WHigh) | High | Very High |
| | | 0-10 | 7-17 | 14-24 | 21-31 | 28-37 | 35-43 | 40-50 |

**Phase 4- Security Testing:** It includes unit testing, acceptance testing, fuzz testing and penetration testing.

**Phase 5- Secure Deployment:** Here deployment of current iteration takes place and with that planning for next iteration starts.

As explained above ASF has various phases incorporating security into every stage of software development. Here, we are extending the work of ASF by providing a fuzzy approach in phase 2 that comprises threat prioritization based on risk during threat modeling. This approach is compatible with practices of agile processes and inherits benefits of security engineering activities in the form of risk assessment and threat prioritization. We will discuss our fuzzy approach in detail in next sections.

## 4. Proposed Fuzzy Logic Implementation

In this approach, a fuzzy logic based technique is designed using fuzzy inference system to determine the risk rate using parameters of DREAD model associated with each threat. Fuzzy inference system is a computer paradigm implying a collection of fuzzy membership function, rules and reasoning. There are three common inference systems known. These are Mamdani Fuzzy models, Sugeno Fuzzy Models, Tsukamoto Fuzzy models. In our approach we are using Mamdani Fuzzy model as it is best suitable to adapt our approach. This fuzzy logic approach proceeds in several steps as shown below.

### 4.1 Defining Input/ Output Variables

For assessing risk level, a risk rating model such as Microsoft's DREAD model can be used, that defines input variables as damage potential, reproducibility of attack, exploitability of the vulnerability, number of affected users and discoverability of vulnerability. Output variable determines risk level associated with a threat. Initially this crisp value is a real number according to universe of scope. Value obtained for each of the input variable is defined by fuzzy number using suitable fuzzy sets.

### 4.2 Designing and fuzzifying linguistic variables

Like crisp input values take on numeric values, in the same way linguistic variables have linguistic values taken as words in the fuzzy logic. In fuzzy logic, for each input variable fuzzy sets are defined as linguistic variables are divided into certain categories as shown in Table1. In each case input parameters range from 0 to 10 while for output range is from 0 to 50.For fuzzification of input parameters, we will define fuzzy membership value for each of the sets using a set diagram called as fuzzy membership curve that graphically defines each of the linguistic value and defines the way in which each point in the input space is mapped to a membership value between 0 and 1. There are several types of membership function. Here we are using triangular membership function (special case of trapezoidal membership function) since they are well suited for modeling and designing.

Table 2: Comparison showing Risk level using conventional approach and Fuzzy Approach

| Threat | D | R | E | A | D | Total | Average | Rating | Fuzzy Risk value | Fuzzy Risk level |
|--------|---|---|---|---|---|-------|---------|--------|------------------|------------------|
| Blind SQL Injection | 9 | 6 | 8 | 9 | 6 | **38** | 7.60 | High | **39** | High (6<sup>th</sup> category) |
| Login Page SQL Injection | 9 | 6 | 8 | 9 | 6 | **38** | 7.60 | High | **39** | High (6<sup>th</sup> category) |
| Unencrypted login request | 6 | 4 | 6 | 5 | 5 | **26** | 5.2 | Medium | **32.5** | Somewhat High (5<sup>th</sup> category) |
| Application Error | 2 | 1 | 3 | 2 | 3 | **11** | 2.2 | Low | **19** | Somewhat Low (3<sup>rd</sup> category) |
| Inadequate account lockout | 2 | 1 | 3 | 2 | 3 | **11** | 2.2 | Low | **19** | Somewhat Low (3<sup>rd</sup> category) |
| Permanent cookie contains sensitive session information | 2 | 1 | 3 | 2 | 3 | **11** | 2.2 | Low | **19** | Somewhat Low (3<sup>rd</sup> category) |
| Session information not updated | 2 | 1 | 3 | 2 | 3 | **11** | 2.2 | Low | **19** | Somewhat Low (3<sup>rd</sup> category) |
| Unencrypted password Parameter | 2 | 1 | 3 | 2 | 3 | **11** | 2.2 | Low | **19** | Somewhat Low (3<sup>rd</sup> category) |
| Unencrypted viewstate Parameter | 2 | 1 | 3 | 2 | 3 | **11** | 2.2 | Low | **19** | Somewhat Low (3<sup>rd</sup> category) |

## 4.3 Development of Fuzzy rule base

If-then rules are framed to reflect relationship between input variables taken as antecedents and output variable taken as consequent of the fuzzy rules. Given input variables specified as 'IF' part of rule and output variable (fuzzy risk level) is taken as 'THEN' part of rule. In our approach there are multiple input variables. That's why we are using AND operator here for mapping of five inputs to one output. Rules designed in the rule base of FIS can be represented in general as

IF (Linguistic Variable1 IS Linguistic Value1) AND (Linguistic Variable2 IS Linguistic Value2) THEN (Linguistic Variable3 IS Linguistic Value3)

## 4.4 Aggregate all outputs

During aggregation outputs of each rule are unified. Here, the input for the aggregation process is truncated output fuzzy sets returned by the implication process for each rule. The output of the aggregation process is the combined output fuzzy set. Here we consider method which computes maximum of each rule's output set.

## 4.5 Defuzzification

It converts the fuzzy output of the inference engine to crisp single output value. Five commonly used defuzzifying methods are Centroid of area (COA), Bisector of area (BOA), Mean of maximum (MOM), Smallest of maximum (SOM), Largest of maximum (LOM). From these methods we are using COA. The input for the defuzzification is a fuzzy set and the output of the process is a value obtained by using a defuzzification method.

After getting risk level, priority for threats can be specified. It can be further used for managing threats by risk acceptance, risk transference, risk removal and for the highest value designer must go for risk mitigation. This is beyond the scope of this paper.

## 5. Case Study

In this section we briefly describe the case study that explores the ramifications of our proposed fuzzy approach by applying it to existing results of a conventional DREAD model technique. We have implemented our approach using Matlab.

Keeping focus on our goal we first provide an overview of our sample data which calculates risk rating using conventional DREAD model on Geospatial Weather Information System (GWIS). GWIS is a web based tool used for performing various operations of weather climatic data. It integrates the weather related information from different available sources and organizes the data in structural GWIS format as explained in [6]. In this paper [6] computation of risk level has been done using conventional DREAD model as shown in Table 2.

For our case study we consider same values of threats and DREAD attributes so that we can compare consequences of risk rating from conventional DREAD model approach with that of our fuzzy approach. Initially using FIS editor of matlab fuzzy toolbox we can define input - output names and also specify different methods and model to be used throughout our implementation as shown in Fig.2. The process of defining linguistic values and their respective ranges is shown in Table 1 as required for GWIS. Then using Membership function editor we represent membership function of each input parameter corresponding to its range, as given in Fig. 3 and 4 for input parameter damage potential and output Risk level. The representations for other input variables can be done in similar way.

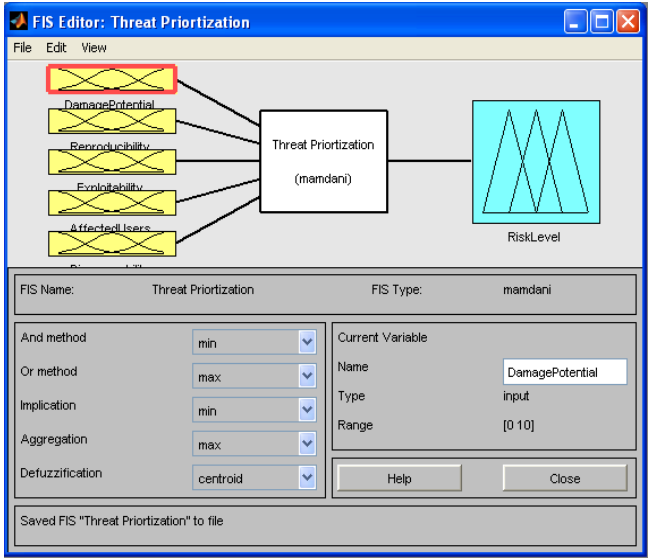

Fig. 2 Representing mapping of inputs to the inference system type and then to the output using FIS editor

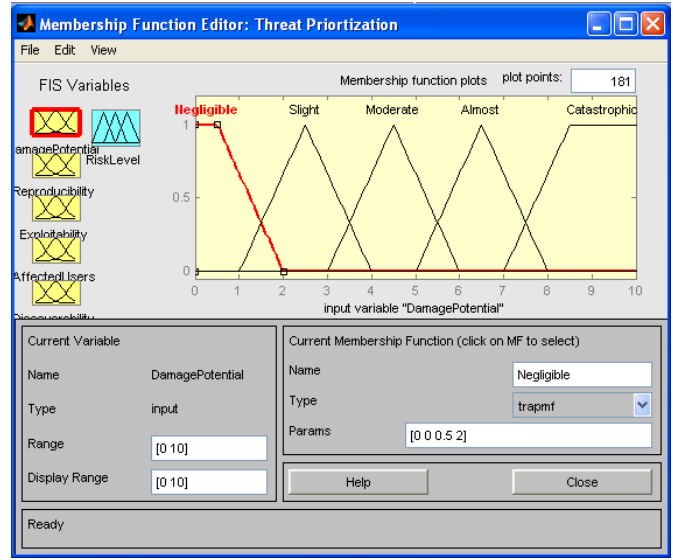

Fig. 3 Membership function editor for Damage Potential

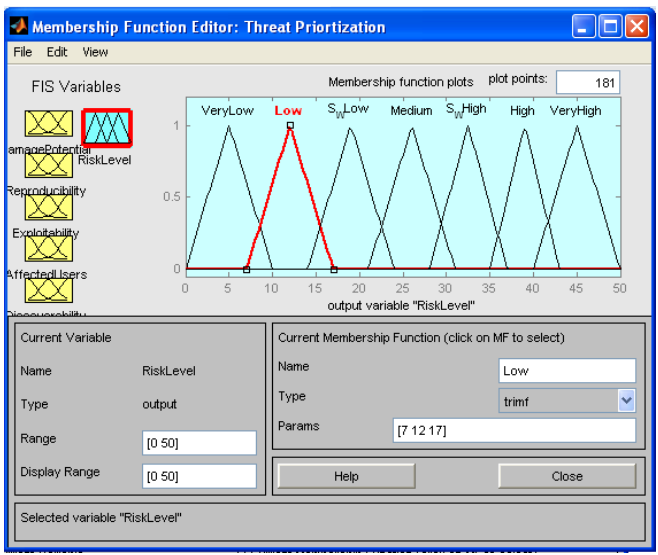

Fig. 4 Membership function editor for Output Risk Level

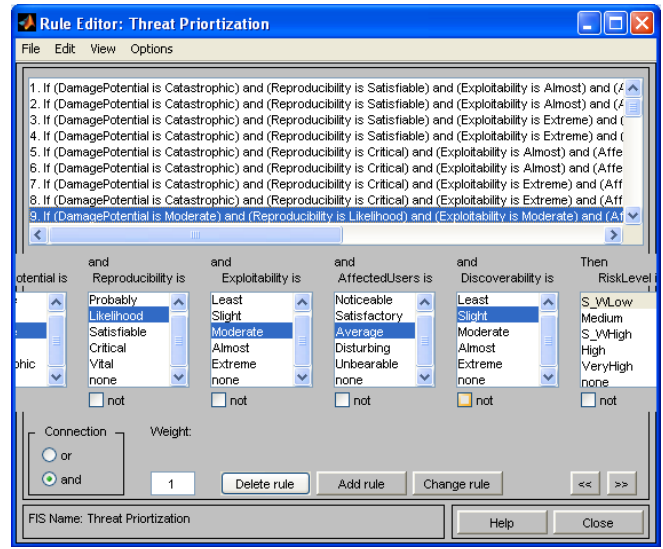

Fig. 5 Rule Editor

More the number of intervals more will be the resolution and preciseness but it is at the cost of computational complexity. Therefore, to balance between the two we are using 5 categories for each linguistic variable, which results into 1225 rules if we consider all input parameters for rule framing. Just few rules may contribute in actual evaluation process based on the input parameter values. However, for a single threat more than one fuzzy rule may apply.

Here, we are considering Blind SQL injection threat from Table 2 having values D=9, R=6, E=8, A=9, D=6. In this case, 8 rules are developed whose implementation has been done using fuzzy conditional statement. Some of them are shown below:

1. If (Damage Potential is Catastrophic) AND (Reproducibility is Satisfiable) AND (Exploitability is Almost) AND (Affected Users is Unbearable) AND (Discoverability is Moderate) then (Risk Level is S_WHigh).

2. If (Damage Potential is Catastrophic) AND (Reproducibility is Satisfiable) AND (Exploitability is Almost) AND (Affected Users is Unbearable) AND (Discoverability is Almost) then (Risk Level is High).

----------------------
----------------------

----------------------
----------------------

7. If (Damage Potential is Catastrophic) AND (Reproducibility is Critical) AND (Exploitability is Extreme) AND (Affected Users is Unbearable) AND (Discoverability is Moderate) then (Risk Level is High).

8. If (Damage Potential is Catastrophic) AND (Reproducibility is Critical) AND (Exploitability is Extreme) AND (Affected Users is Unbearable) AND (Discoverability is Almost) then (Risk Level is VeryHigh).

Now, we can define fuzzy membership value for each set using membership function diagram as shown in Fig. 3, 4. The fuzzification output can be represented as:

1 Damage Potential at 9 have $\mu_{DP}$ = 1 f or 'Catastrophic' membership function AND
Reproducibility at 6 have $\mu_R$ = 0.3 for 'Critical' membership function AND
Exploitability at 8 have $\mu_E$ = 0.7 for 'Almost' membership function AND
Affected Users at 9 have $\mu_{AU}$ = 1 for 'unbearable' membership function AND
Discoverability at 6 have $\mu_D$ = .65 for 'Moderate' membership function.

2 Damage Potential at 9 have $\mu_{DP}$ = 1 f or 'Catastrophic' membership function AND
Reproducibility at 6 have $\mu_R$ = 0.1 for 'Critical' membership function AND
Exploitability at 8 have $\mu_E$ = 0.7 for 'Almost' membership function AND
Affected Users at 9 have $\mu_{AU}$ = 1 for 'unbearable' membership function AND
Discoverability at 6 have $\mu_D$ = .65 for 'Moderate' membership function.

----------------------
----------------------

----------------------
----------------------

8 Damage Potential at 9 have $\mu_{DP}$ = 1 f or 'Catastrophic' membership function AND
Reproducibility at 6 have $\mu_R$ = 0.3 for 'Critical' membership function AND
Exploitability at 8 have $\mu_E$ = 0.7 for 'Almost' membership function AND
Affected Users at 9 have $\mu_{AU}$ = 1 for 'unbearable' membership function AND

Discoverability at 6 have to $\mu_D$ = .25 for 'Almost' membership function.

After deriving these rules, AND fuzzy operator is used in between the antecedents of the rules. Therefore, minimum of all membership functions on antecedents side of a rule is calculated, which gives a single membership value. Using this membership value of antecedent side, the consequent side of risk is evaluated. Here, 8 rules are applied for Blind SQL injection threat and each applicable rule contributes a vote of membership. Weight applied to each rule is 1. Overall, 48 rules are used in our case study for the threats we considered as shown in Fig. 5. Finally, from rule viewer represented in Fig. 6, evaluation of output fuzzy risk level is performed for the threats faced by GWIS web tool. The computation of each threat is considered separately by giving input values of each parameter from Table 2. The last two columns of Table 2 represent fuzzy risk value we get by our approach and fuzzy rating we can assign to it

Now we can compare the results obtained for risk level evaluation using conventional DREAD model and our fuzzy logic approach. We have discovered from Table 2 that risk level obtained from our fuzzy approach is significantly more than that of existing results of conventional approach. Our study suggests that this is due to fact that conventional approach has sharp cutoffs for variables. That is, the member certainly belongs to a set or not which inferences that precision level required for this approach is too high. For example, suppose in given Table 2 when a variable Total lies between the interval of 20 to 30 then risk rate is medium. Now we consider a given threat of unencrypted login request which rates medium if total is exactly 20, 30 or somewhere in between 20 to 30. But instead of that if its total is 19 then it arrives in low rating and if 31 then high rating. It shows that just by a small change in some attribute of DREAD model class of risk rating gets deviated. Although values assigned to attributes are subjective quantities and can vary from person to person by 1 to 2 degrees. Introduction of such sharp boundaries is undesirable as risk is vague and uncertain term. Our fuzzy approach smoothes the edges and here transition takes place gradually. In fuzzy logic, a variable can be a member of multiple fuzzy sets having different degree of membership for each set. Therefore, it includes all the values for an object to which they belong in the fuzzy sets. But in traditional modeling only true values are considered and it ignores some values which are partially true. Thus here risk value arises less than that of fuzzy risk value. Also more than one fuzzy rule is applied as each applicable rule contributes a vote of membership. This all concludes that our approach is more effective for unpredictable risks and perhaps easier to implement.

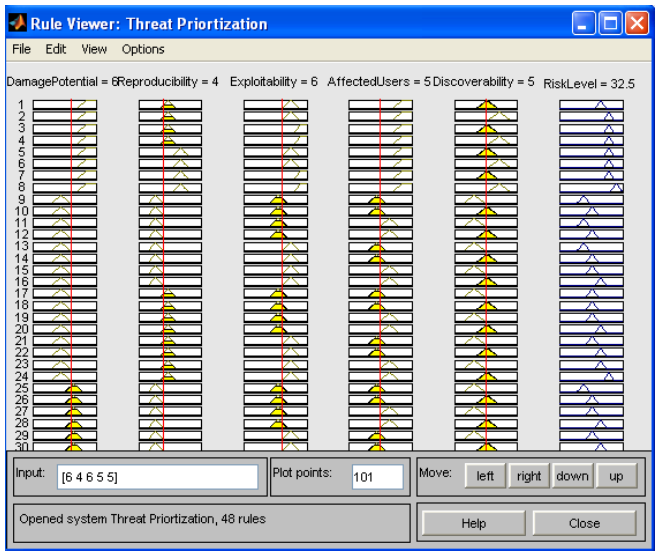

Fig. 6 Rule Viewer computing Ouput by combining rules

Furthermore, after careful analysis we can say that fuzzy approach is more efficient, flexible and having more tolerance for imprecise data and ambiguities like that of risk. For instance, as seen from Table 2 that rating for unencrypted login request for a web application GWIS is medium but after referring [16], we can see that same threat for a web application using DREAD itself gets high rating. That is, conventional DREAD model can lead to variations. Our new approach has been tested and verified for same values and rating results from it is 'somewhat high'. Thus using fuzzy logic, we can capture the notion that risk of unencrypted login request is high to some degree (i.e somewhat high) and not medium as in conventional method, although not as high as that of other risks posing severe inevitable problems. Results represented in Table 2 for our fuzzy approach can describe various levels to be considered more or less based on severity rather than yes or no description. It also has the advantage of being simple to implement and qualitative in nature as it provides high level of abstractions by using linguistic labels. However, in conventional model, risk analysis is quantitative and for human beings mathematical formulation using numbers is more difficult.

Our approach provides more resistance to the problems posed by uncertainties always associated with risks and also able to handle ambiguities effectively for threat prioritization. Therefore more severe threats can be addressed immediately during earlier stages. This prevents security engineers from facing complex and costly problems in later stages.

## 6. Conclusion and Future Work

For current software system it is worth saying that, sound security practices are enlightening but are not completely life savers. Our work refines the way of enforcing security in the well established Agile Security Framework. One of the major aspects perceived by ASF is risk assessment and threat prioritization. In common practice, the risk associated with threats are rated using DREAD model. In the given paper we are recommending a fuzzy approach for carrying out threat prioritization based on risk ranking. Here transition from member to non member is gradual rather than abrupt. It can easily deal with uncertain, ambiguous and vague risks as it a llows description of concepts by eliminating sharp boundaries between members of class from non members. So prioritization of threats based on risk level is more efficient, effective and simple. Use of fuzzy sets and their linguistic values during risk analysis make valuable contributions as they are able to represent abstractions, which is natural for human being instead of numbers. Although we have implemented our fuzzy approach in agile environment but it can also work perfectly in traditional development process. Here we have made an attempt to improve the concept of prioritizing threats but there is a great deal of work ahead for providing a perfect secure system. Threats selected for mitigation considers risk level and cost of recovery. Extensive security measures providing fewer benefits can increase the cost of system. This factor definitely provides a direction for future work.